\documentclass[12pt,preprint]{aastex}

\newcommand{\kms}{\rm ~km~s^{-1}}

\newcommand{\dyn}{\rm ~dyn~cm^{-2}}

\begin{document}

\title{PULSAR WIND NEBULAE WITH THICK TOROIDAL STRUCTURE}
\author{
Roger A. Chevalier\altaffilmark{1} and
Stephen P. Reynolds\altaffilmark{2}
}
\altaffiltext{1}{Department of Astronomy, University of Virginia, P.O. 
Box 400325,
Charlottesville, VA 22904-4325; rac5x@virginia.edu}
\altaffiltext{2}{Physics Department, North Carolina State University,
Raleigh, NC 27695; reynolds@ncsu.edu}

\begin{abstract}
We investigate a class of pulsar wind nebulae that show synchrotron emission
from a thick toroidal structure.
The best studied such object is
the small radio and X-ray nebula around the Vela pulsar, which can be interpreted as the result
of interaction of a mildly supersonic inward flow with the recent pulsar wind.
Such a flow near the center of a supernova remnant can be produced
in a transient phase when the reverse shock reaches the center of the remnant.
Other nebulae with a thick toroidal structure are G106.6+2.9 and G76.9+1.0.
Their structure contrasts with young pulsar nebulae like the Crab Nebula and 3C 38, which show a
more chaotic, filamentary structure in the synchrotron emission.
In both situations, a torus-jet structure is present where the pulsar wind passes through
a termination shock, indicating the flow is initially toroidal.
We suggest that the difference is due to the Rayleigh-Taylor instability that operates when the 
outer boundary of the nebula is accelerating into freely expanding supernova ejecta.
The instability gives rise to mixing in the Crab and related objects, but is not present
in the nebulae  with thick toroidal regions.

\end{abstract}


\section{INTRODUCTION}

The structure of a pulsar wind nebula is determined by the interaction of
the relativistic pulsar wind with its surroundings, which are initially the
parent supernova and later the interstellar medium.
Pulsar nebulae are best observed by their synchrotron radio and X-ray emission, and
three classes of nebulae can be characterized \citep{gaensler06}.
The first, typified by the Crab Nebula \citep{hester08}, shows a small toroidal structure
close to the pulsar, and a more symmetric, larger scale structure.
The outer boundary is formed where the shocked pulsar wind expands into
the freely expanding gas of the parent supernova.
The second, typified by Vela X,  is formed after the reverse shock wave
from the interaction with the interstellar medium moves back to the pulsar nebula \citep{blondin01}.
This action displaces the pulsar nebula from the pulsar, leaving a `relic' nebula.
The third, typified by the Mouse \citep{gaensler04}, shows a trail of emission as the pulsar
moves rapidly through the interstellar medium.

Here, we consider members of another class of pulsar nebula that is characterized by a thick
toroidal emission region.
In Section 2, we treat various members of this class, with special attention to
the recently formed nebula surrounding the Vela pulsar.
The distinguishing attributes of this class of pulsar nebula are discussed in Section 3.

\section{PULSAR NEBULAE WITH THICK TOROIDAL EMISSION REGIONS}

\subsection{The Vela Nebula}

The best studied example of a remnant with a likely asymmetric reverse shock is the Vela
remnant, where a strong case can be made that a reverse shock from the N
side of the remnant has displaced the original pulsar 
nebula (Vela X) from the pulsar \citep{blondin01}.
While Vela X is now a relic nebula, the Vela pulsar has created a new nebula in its vicinity.
Deep radio images of the region around the Vela pulsar show evidence for a bow wave
\citep[Figs. 9 and 10 of][Fig. 1]{dodson03a}.
The orientation of the bow wave is consistent with a flow from the N direction;
the fact that the emission is strongest along a NE-SW line can be attributed to the 
orientation of the pulsar equatorial plane with this direction.
Deep X-ray images show similar structure to that seen in the radio \citep{pavlov03,karg04}.
\cite{pavlov03} note that the X-ray jet  to the NW tends to flop to the S, indicating
the presence of a flow in that direction in the shocked pulsar wind.

The presence of a weak bow shock in Vela might seem surprising because the pulsar is
centrally located in the larger remnant and \cite{van04} showed that a moving pulsar
must be at least 68\% of the way to the outer shock in order to be moving supersonically.
However, \cite{van04} assume that the supernova remnant flow is described by the
Sedov solution.
There is a transition period before the Sedov solution is closely
approximated.
Considering a centrally located pulsar, the postshock region of the reverse shock (speed $v_s$)
is characterized by a velocity $0.75v_s$ and sound speed $c_s=(5p/3\rho)^{1/2}$,
with $p=0.75 \rho_0v_s^2$ and $\rho=4\rho_0$ from the shock
conditions, where $\rho_0$ is the preshock density.
The result is  a Mach number ${\cal M}=v/c_s=3/5^{1/2}=1.3$ flow.
If the pulsar has a velocity that carries it away from the center, it is comoving
with the freely expanding gas, so that the properties of the postshock gas
relative to the pulsar are
the same as just described.
If the reverse shock front has recently passed over the Vela pulsar, a
mildly supersonic flow is naturally produced.

The observed radio emission marks the shocked pulsar wind and thus the black line in
Fig. 1 delineates the contact discontinuity (CD) between the shocked pulsar wind
and the shocked surrounding medium.
The half-opening angle of the CD at large distances is $\sim 30^{\circ}$.
The opening angle of the bow shock is expected to be larger \citep[e.g., Fig. 8 of][]{van04}.
In flow past a blunt object, the half-opening angle, $\beta$, of the shock wave
that is generated obeys $\sin \beta=c_s/v=1/\cal M$, where $c_s$ is the gas sound speed,
$v$ is the gas velocity, and $\cal M$ is the Mach number \citep[e.g.,][]{landau}.
The expected ${\cal M}=1.3$ flow thus gives a half-opening angle for the bow shock
of $50^{\circ}$, which is roughly consistent with the observations.
The implication is that there is a mildly supersonic flow past the pulsar nebula.

The direction of pulsar motion is known
\citep[position angle $301^{\circ}$,][]{dodson03b}
 and does not line up with the displacement of Vela X or the bow shock discussed here.
The flow velocity involved in the displacement of Vela X is $v_d=500(r_d/{\rm 5~pc})(t_d/{\rm 2500~yr})^{-1}\kms$,
where $r_d$ is the displacement distance estimated for the center of the relic pulsar
nebula and $t_d$ is the age since the displacement began.
The temperature in the central region of Vela is $\sim 1.2$ keV \citep{lu00}, 
yielding a sound speed of $560\kms$, although there are also cooler components.
The flow velocities associated with the reverse shock are apparently much larger
than the projected velocity of the Vela pulsar, $61\kms$ \citep{dodson03b}, so that
the pulsar velocity does not play a significant role in the orientation of the flow
past the pulsar, in agreement with observations.
If the surroundings of the pulsar have been cleared so that the pulsar interacts
directly with the thermal gas, a mildly supersonic flow past the pulsar is expected.

Some properties of the Vela pulsar nebula are given in Table 1.
The nebular pressure $P_n$ at the termination shock (assumed to be the torus radius $R_{torus}$)
is given by pressure balance $P_{n}=\dot E/(4 \pi R_{torus}^2 c)=2\times 10^{-9}\dyn$
for the parameters  in Table 1, where $\dot E$ is the spindown power of the pulsar.
If half of this pressure is supplied by the magnetic field, we have $B=1.6\times 10^{-4}$ G.
Assuming $\gamma=4/3$ for the shocked pulsar wind and a uniform pressure, the energy
(within the nebular radius $R_{radio}$) is $E_{neb}=3.6\times 10^{45}$ ergs.
The minimum energy, $E_{min}$, required to produce the synchrotron emission from the Vela nebula can be estimated.
The total flux of the radio nebula is $ 1$ Jy and has a flat spectrum between
1.4 and 8.5 GHz \citep{dodson03a}.
Extrapolating the low energy X-ray emission, with $\alpha=-0.66$
where $F_{\nu}\propto \nu^{\alpha}$ \citep{mangano05},
yields a break frequency of $3\times 10^{10}$ Hz.
Equation (46) of \cite{chevalier05} for the minimum energy
then yields $E_{min}=4\times 10^{44}$ ergs.
The energy estimated above is $E_{neb}\sim10$ times the minimum.
The time for the pulsar spindown to produce this energy is $E_{neb}/\dot E= 17$ yr.
Radiative losses are small and adiabatic expansion losses are not expected to change
this estimate by a large factor ($\la 2$), so the age of the particles in the
radio nebula is relatively small.
Older particles are swept downstream in the SE direction, as can be seen in deep
radio and X-ray images \citep{dodson03a,karg04}.

As mentioned above, the spectrum of the total radio flux is flat, or slowly declining,
which is typical of pulsar wind nebulae.
However, \cite{dodson03a} find large (several 0.1's) variation in $\alpha$ over the
nebula when examined over the separate frequency ranges.
Also surprising is that in some places $\alpha>1/3$, which is not expected
for optically thin synchrotron radiation.
\cite{bieten91} measured $\alpha=-0.6$ between 1.4 and 5 GHz for the
bright part of the radio nebula to the NE, which also indicates variations in spectral index.
This situation contrasts with that of the Crab and related nebulae, where spatial
variations in radio spectral index are small.
For the Crab, \cite{bieten97} find that the $\alpha$ is constant to an accuracy of 0.01.
Although the limit on variation is weaker in the case of 3C 58, the spectral index is
also fairly constant \citep{bieten01}.
These properties suggest that there are mixing or diffusive processes going on in
the Crab and 3C 58 that are not present in the Vela nebula.

The morphology of the Vela radio nebula contrasts with the pre-reverse shock pulsar nebulae.
The magnetic field is predominantly toroidal, with the linear polarization
reaching 60\% \citep{dodson03a}.
Filamentary structure with a scale comparable to
the termination shock radius, as seen in younger nebulae, is not apparent.
The radio nebula is smaller than in the pre-reverse shock cases, but
$R_{radio}/R_{torus}=5-6$; the angular size of the structures is comparable
to the case of the Crab Nebula.
We suggest that the absence of structure is due to a combination of 
the facts that the outer boundary is
not subject to the 
Rayleigh-Taylor instability in this case and that only relatively recent effects
of the pulsar wind are observed.
The strong emission in the pulsar equatorial plane can be attributed to
the stronger pulsar power in this direction, as expected in a split monopole
model for the pulsar wind \citep{bogovalov99}.
The dynamics of the nebula depend on a complex 3-dimensional situation because
the rotation axis is not aligned with the direction of the surrounding flow.
The fact that a jet-torus structure is present around the pulsar suggests that
the magnetization parameter
$\sigma\approx 0.1$ in the pulsar wind \citep{buccian11} and that the magnetic
field reaches equipartition strength in the downstream flow.
Once the magnetic field reaches this strength in a bow shock situation, it
tends to increase the size of the wind nebula between the termination shock
and the contact discontinuity \citep{buccian05a}, so the contact discontinuity
is at several times the termination shock radius,
as observed in the Vela nebula.

Another indication of significant magnetic forces in the nebula  is the central position
of the pulsar in the inner X-ray nebula \citep[Fig. 1,][]{ng04}.
In a hydrodynamic model in which the pressure is constant between the
contact discontinuity and the pulsar wind termination shock, the ratio of the
termination shock radius on the upstream side to that
on the downstream side is $1/\cal M$
\citep{van03}.
Assuming that the inner X-ray nebula represents the termination shock and ${\cal M}\approx 1.3$, 
the  displacement expected in the hydrodynamic model is not present.
However, significant pressure forces from a toroidal magnetic field give a decreasing
pressure with radius and the pressure gradient can
be larger than the pressure differential from the bow shock estimated here \citep{kennel84a,emmering87}.

We have examined simple models for the projected radio emission in which there is 
uniform emission from a toroidal region with an inner and outer radius, and an 
opening angle.
Fig. 2 shows a model that resembles the radio structure \citep{dodson03a}:
the outer radius is 5 times the inner radius, which is assumed to be at the
position of the inner X-ray torus,  the half-angle of the disk is 
$35^{\circ}$, and the angle between
the line of sight and the axis of the torus is $\zeta=50^{\circ}$.
The emissivity is assumed to be uniform and isotropic, which would be appropriate
for a region with a tangled magnetic field.
However, the observed radio polarization at 5 GHz \citep{dodson03a} is the maximum expected for a flat flux spectrum,
suggesting that the magnetic field is  ordered.
Emission in a purely toroidal field would enhance the emission along the minor axis
compared to the major axis and thus reduce the agreement with the observed
morphology.
Models for the synchrotron emission from the central region of the Crab 
based on relativistic MHD simulations show reasonable
agreement with the observations \citep{buccian05b,delzanna06}.
The outer parts of
ring-like structures are enhanced when the magnetic field is taken to
be isotropic as opposed to toroidal \citep[][their Figs. 3 and 4]{delzanna06}.
\cite{delzanna06} also find the effects of the mildly relativistic flow are significant.
In our case, the radial velocity drops beyond the termination shock 
so that relativistic effects are not important.

The observed emission from the Vela nebula shows a high surface
brightness to the NE \citep{dodson03a}, presumably due to the low Mach number flow.
The flow is probably also responsible for the higher surface brightness to the NW along
the minor axis, and possibly for the brighter X-ray ``outer jet'' to the NW even though its
motion is expected to be directed away from us based on the brightness distributions
of the inner tori \citep{pavlov03,ng04}.
The angle $\zeta$ in our model is somewhat less than that found by \cite{ng04} for the
inner toroidal X-ray structure around the pulsar ($\zeta=63.6^{\circ}$); again, the
difference may be related to the bow shock flow.
Three-dimensional MHD simulations will ultimately be needed to model the emission.

The X-ray emission is observed to extend out to the same place
as the radio \citep{karg04}.
The age of the nebula and the magnetic field strength indicate a synchrotron
cooling break in the X-ray range and
\cite{mangano05} find an apparent cooling break in the spectrum at 12.5 keV.
This would explain the similar extent at radio and X-ray wavelengths.

\subsection{G106.6+2.9}

Another nebula with similar properties to the Vela nebula is G106.6+2.9 
(also known as the Boomerang), associated with J2229+6114 (Table 1).
Unfortunately, the distance is poorly known, with estimates of
800 pc based on low radio depolarization and interaction with neutral
gas \citep{kothes01}, 3 kpc based on X-ray absorption \citep{halpern01},
and 7.5 kpc based on pulsar dispersion measure \citep{abdo09}.
We scale the distance, $d$, with $d_3=d/(3{\rm~kpc})$.
The Boomerang has considerable radio emission on one side and \cite{kothes06}
identified this structure as the result of an asymmetric reverse shock, although
a larger scale thermal remnant is not observed around the nebula.
In this case, the pulsar $\dot E$ and $R_{torus}$ suggest $P_{n}=3.3\times 10^{-10}d_3^{-2}\dyn$.
The magnetic field and the radio nebula energy, computed as for Vela, are
$B=6.5\times 10^{-5}d_3^{-1}$ G and $E_{neb}= 2.7 \times 10^{47}d_3$ ergs.
The minimum energy in the radio nebula is $2.4\times 10^{46}d_3^{17/7}$ ergs.
The time for the pulsar to produce the nebular energy is again relatively short, $\sim 390d_3$ yr.

The radio nebula is highly polarized, up to 50\%, indicating a toroidal field component \citep{kothes06}.
The radio spectrum shows $\alpha=-0.11\pm 0.05$ below 4.3 GHz and $\alpha=-0.59\pm 0.09$ above,
which \cite{kothes06} interpret as a synchrotron cooling break; they note that there is a steeper
spectrum farther from the pulsar.
To obtain such a low cooling break frequency, \cite{kothes06} need a magnetic field of 2.6 mG and
particle lifetime of 3900 yr in a model with $d_3=0.27$.
We find that this field would produce too high a pressure in the nebula and the age is longer
than expected, so that the spectral break is likely to be intrinsic to the injection.

\cite{kothes06} propose that the variations in rotation measure across the remnant are
intrinsic to the remnant because of the apparent symmetry about an axis defined by the
polarized emission.
For the magnetic field we have derived,  an electron density $n_e\ga 1/f$ cm$^{-3}$, where
$f$ is the filling factor would be needed in the remnant to produce the rotation measure.
In our view, $n_e$ is much lower than this value because the pulsar wind bubble has
smoothly expanded into the surrounding medium, so the apparent alignment of the
rotation measure pattern is a coincidence.

\subsection{G76.9+1.0}

The remnant G76.9+1.0  has a similar radio
morphology to the Vela nebula,
and {\it Chandra} observations show that the radio lobes have a similar
orientation compared to the axis of the pulsar rotation \citep{arz11}.
However, it differs from the Vela nebula in 2 important ways:
it is much larger ($29\times 35$ pc at a distance of 10 kpc) and the large scale nebula
has symmetry about the pulsar.
These properties indicate that the pulsar nebula has not been affected
by the reverse shock wave.
We suggest that the inward mixing of thermal filaments has not
been so efficient in this case so that a toroidal emission region has
a chance to build up, as in the case of the Vela nebula.
The instabilities are expected to be strongest during the constant
$\dot E$ phase of the pulsar evolution.
Once a pulsar is no longer adding substantially to the energy of the
remnant, the acceleration ceases.  This might be expected in a 
large pulsar nebula where a low density surroundings has delayed
the action of the reverse shock wave.

\section{DISCUSSION}

Pulsars are expected to preferentially inject their wind power close to the
equatorial plane and models that allow for such power injection
\citep{komissarov03,delzanna04} have been successful in reproducing many
of the properties of the torii observed immediately surrounding the pulsars in
young wind nebulae.
However, on a larger scale (several inner torus radii), the young nebulae such as the
Crab and 3C 58 do not show the toroidal structure.
Here we have investigated the evolutionary status of several  pulsar
nebulae, best exemplified by the newly formed Vela nebula, that do show
a thick toroidal structure.
We suggest that the structure is not present in the Crab and related objects
because of Rayleigh-Taylor instabilities at the interface between the expanding
supernova gas and the accelerating pulsar nebula \citep{chev77,hester08}, although
there remain uncertainties in the instability when the light fluid is magnetized
\citep{buccian04,stone07}.
The instabilities give rise to mixing and diffusion of the relativistic particles.
In the cases discussed here, the instability is absent either because the
outer boundary is determined by a somewhat asymmetric outer pressure
or the pulsar power has declined so the outer boundary is no longer accelerated.
The age at this phase is $\sim 10^4$ yr, within a factor of a few.
In the case of an asymmetric outer pressure, a relic pulsar nebula may also be
present, as in the case of the Vela nebula.
At later times, the pulsar velocity leads to a nebular trail to the pulsar.

Multidimensional simulations are needed to confirm the scenario proposed here. 
In their numerical simulations of pulsar nebulae, \cite{camus09} find that waves and vortices
in the larger nebula feed back on the structure at the termination shock,
which in turn generates more structure in the nebula; they note that it
is impossible to distinguish between the cause and the effect.
In addition, some thermal gas from the supernova is entrained in the unstable
region, explaining the optical filaments seen in association with nonthermal
filaments in the Crab and 3C 58.
These were 2-dimensional simulations, but did not impose symmetry about the
equatorial plane.
More investigation of the instabilities and 3-dimensional simulations are
ultimately needed \citep[e.g.,][]{mizuno11}.

\acknowledgments
We are grateful to Richard Dodson for permission to use the image shown in Fig. 1,
and to the referee for helpful comments.
This research was supported in part by
NASA grant NNX07AG78G.


\clearpage

\begin{center}
\noindent{{Table 1.} Properties of Pulsars and their Nebulae}

\begin{tabular}{ccccccccc}
\hline
PSR & Supernova & Distance & $\dot E/I_{45}$ & $\tau_{char}$&$R_{torus}$ & $R_{radio}$ & Refs.   \\
     &    Remnant  & (kpc)  & (ergs s$^{-1}$)&(yr) & (pc) & (pc)  &  \\    
\hline
B0833--45 &  Vela &  0.29 &  $6.9\times 10^{36}$   & $1.1\times 10^4$ &  0.030   & 0.17  & 1,2,3   \\

J2229+6114 &  G106.6+2.9  & $3d_3$  &  $2.2\times 10^{37}$  & $1.05\times 10^4$ & $0.13d_3$ & $1.3d_3$  &  4,3  \\
\hline
\end{tabular}
\end{center}
\noindent{References.--  (1) \cite{dodson03b}; (2) \cite{dodson03a}; (3)  \cite{ng04}; (4) \cite{kothes06}}

\clearpage
\begin{figure}[!hbtp]   
\epsscale{.80}
\includegraphics[totalheight=0.7\textheight, angle=180]{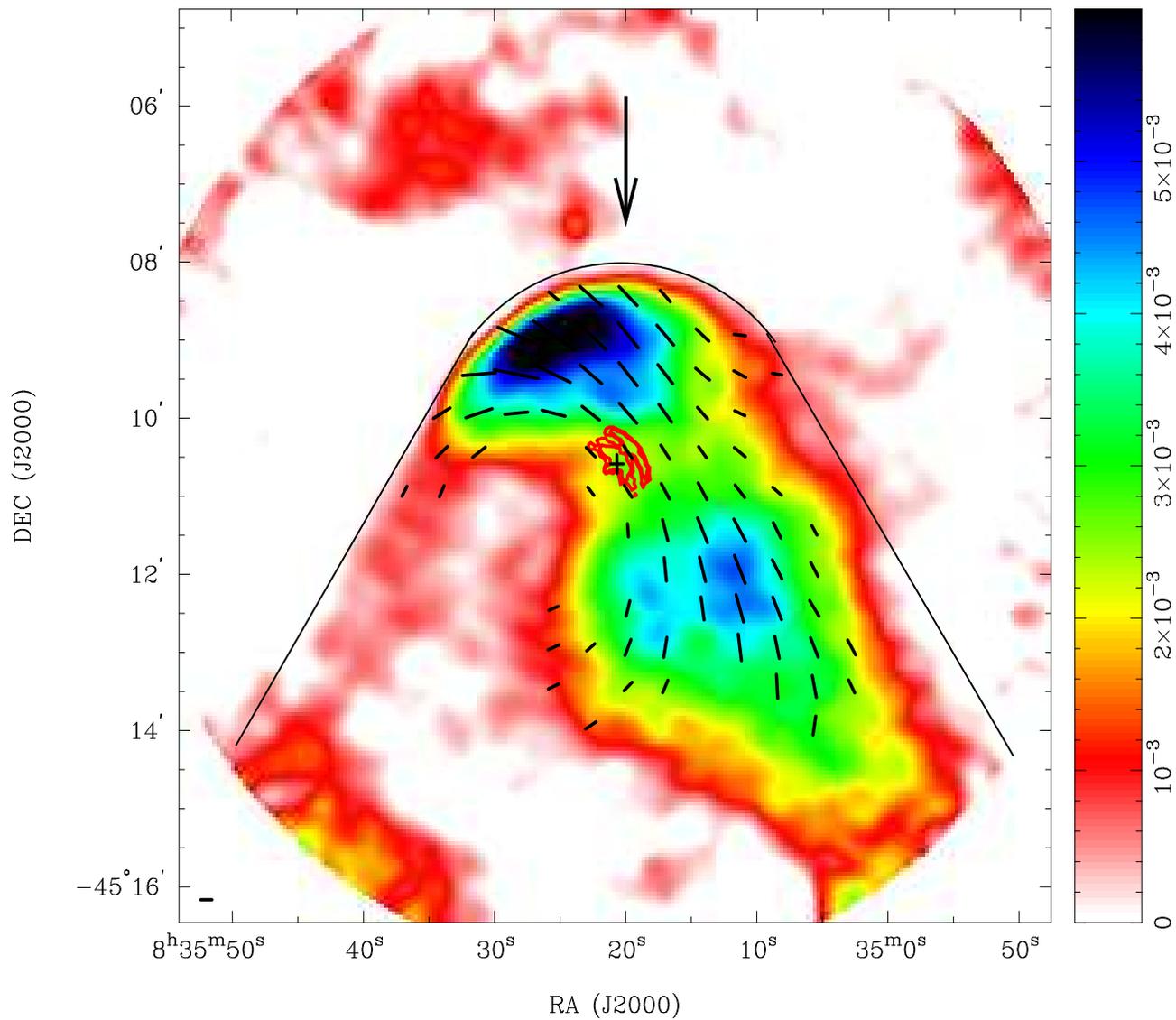}
\caption{The image, reproduced from Fig.\ 10 of \cite{dodson03a}, shows the Vela nebula at 5 GHz, with derotated magnetic field lines
 and Chandra X-ray contours overlaid.
The cross marks the pulsar position.
The solid black line is the suggested contact discontinuity between the shocked pulsar wind and gas that has passed through the bow shock.
The arrow indicates the direction of the mildly supersonic flow.
}
\end{figure}

\clearpage
\begin{figure}[!hbtp]   
\epsscale{.80}
\plotone{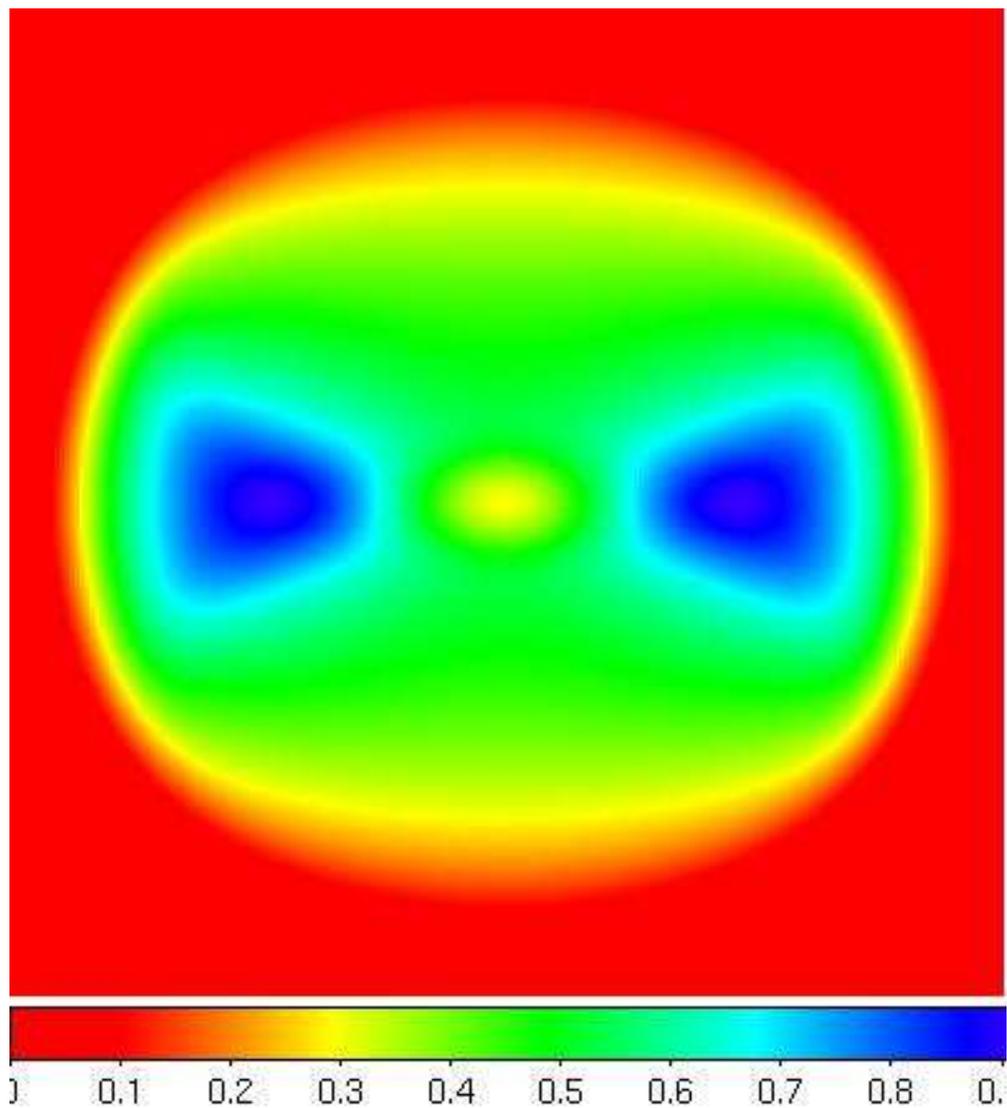}
\caption{Model of a toroidal  region with uniform emissivity seen in projection
with the axis tilted 50$^{\circ}$ to the line of sight.
The disk half-angle is 35$^{\circ}$ and the maximum radius is 5 times the inner radius.
The color scale is linear.
}
\end{figure}

\end{document}